\documentclass[aps,prb,twocolumn,superscriptaddress,floatfix]{revtex4-1}

\usepackage{graphicx,graphics}
\usepackage{dcolumn}
\usepackage{amsmath,amssymb,amsfonts}
\usepackage{latexsym,verbatim}
\usepackage{bm}
\usepackage{color}
\usepackage{ulem}
\usepackage[percent]{overpic}
\usepackage[breaklinks=true,colorlinks,citecolor=blue,linkcolor=blue,urlcolor=blue]{hyperref}

\begin{document}
\title{Nonlocal superconducting correlations in graphene in the quantum Hall regime}
\author{Michael Beconcini}
\email{michael.beconcini@sns.it}
\affiliation{NEST, Scuola Normale Superiore, I-56126 Pisa,~Italy}
\author{Marco Polini}
\affiliation{Istituto Italiano di Tecnologia, Graphene Labs, Via Morego 30, I-16163 Genova,~Italy}
\author{Fabio Taddei}
\affiliation{NEST, Istituto Nanoscienze-CNR and Scuola Normale Superiore, I-56126 Pisa,~Italy}
\begin{abstract}
We study Andreev processes and nonlocal transport in a three-terminal graphene-superconductor hybrid system under a quantizing perpendicular magnetic field [G.-H. Lee {\it et al.}, Nature Phys.~{\bf 13}, 693 (2017)]. We find that the amplitude of the crossed Andreev reflection (CAR) processes crucially depends on the orientation of the lattice.
By employing Landauer-B\"{u}ttiker scattering theory, we find that CAR is generally very small for a zigzag edge, while for an armchair edge it can be larger than the normal transmission, thereby resulting in a negative nonlocal resistance. In the case of an armchair edge and with a wide superconducting region (as compared to the superconducting coherence length), CAR exhibits large oscillations as a function of the magnetic field due to interference effects. This results in sign changes of the nonlocal resistance. 
\end{abstract}

\maketitle

{\it Introduction.---}The electron-transport properties of hybrid superconducting two-dimensional (2D) systems in high magnetic fields
have attracted considerable experimental~\cite{takayanagi_physb_1998,moore_prb_1999,eroms_prl_2005,batov_prb_2007} and theoretical~\cite{takagaki_prb_1998,asano_prb_2000,hoppe_prl_2000,giazotto_prb_2005,chtchelkatchev_prb_2007,khaymovich_epl_2010} interest in the last decades. In the quantum Hall (QH) regime, the conduction of charge in a 2D electron gas (2DEG) occurs only along the edges of a Hall bar via 1D chiral edge states.
When a 2DEG/S interface is formed, by placing a superconductor in contact with the Hall bar, pairs of electrons are transferred through the interface via the Andreev reflection process, in which an electron impinging on the interface is backscattered as a hole.
This gives rise to the formation of the so-called Andreev edge states that propagate along the 2DEG/S interface and, in a quasi-classical picture, consist of alternating electron and hole orbits~\cite{hoppe_prl_2000}. For an interface with weak disorder and a small Fermi wavelength mismatch, strong conductance oscillations as a function of magnetic field have been predicted due to interference between the electron and hole parts of the Andreev edge states~\cite{hoppe_prl_2000,giazotto_prb_2005,khaymovich_epl_2010}. 

The quality of the contacts with a superconductor may be improved by using graphene (G) in place of ordinary superconducting 2DEGs. The absence of a band gap ensures low contact resistance and weak scattering at G/S interfaces~\cite{calando_natphys_2015,benshalom_natphys_2015,amet_sci_2016}.
In particular, G encapsulated in hexagonal boron nitride (hBN) exhibits very high mobilities and ballistic transport~\cite{mayorov_prb_2011,wang_sci_2013,beconcini_prb_2016}. These properties, along with the ability to control the filling factor by varying the electronic density with a gate-voltage, make G an ideal platform for exploring Andreev physics in 2D systems\cite{hou_prb_2016,rainis_prb_2009,akhmerov_prl_2007,sun_jpcm_2009}.
Interestingly, Andreev reflection in G nanoribbons (GNRs) is sensitive to the ribbon width and the pseudoparity of quantum states~\cite{rainis_prb_2009} in zero magnetic field. Moreover, in the QH regime, the Andreev scattering processes for the lowest Landau level have been found to depend on the relative angle between the edges where initial and final scattering states propagate~\cite{akhmerov_prl_2007,sun_jpcm_2009}.

Recently, in Ref.~\onlinecite{lee_natphys_2017} transport through a G/S system consisting of a GNR containing a superconducting finger inclusion was experimentally investigated in the QH regime.
A negative nonlocal resistance was measured, between two normal contacts, and claimed to be a direct consequence of the presence of crossed Andreev reflection (CAR--the process by which an electron impinging on one side of the S finger is transmitted as a hole on the opposite side).

\begin{figure}[t]
\centering
\includegraphics[width=0.9\linewidth]{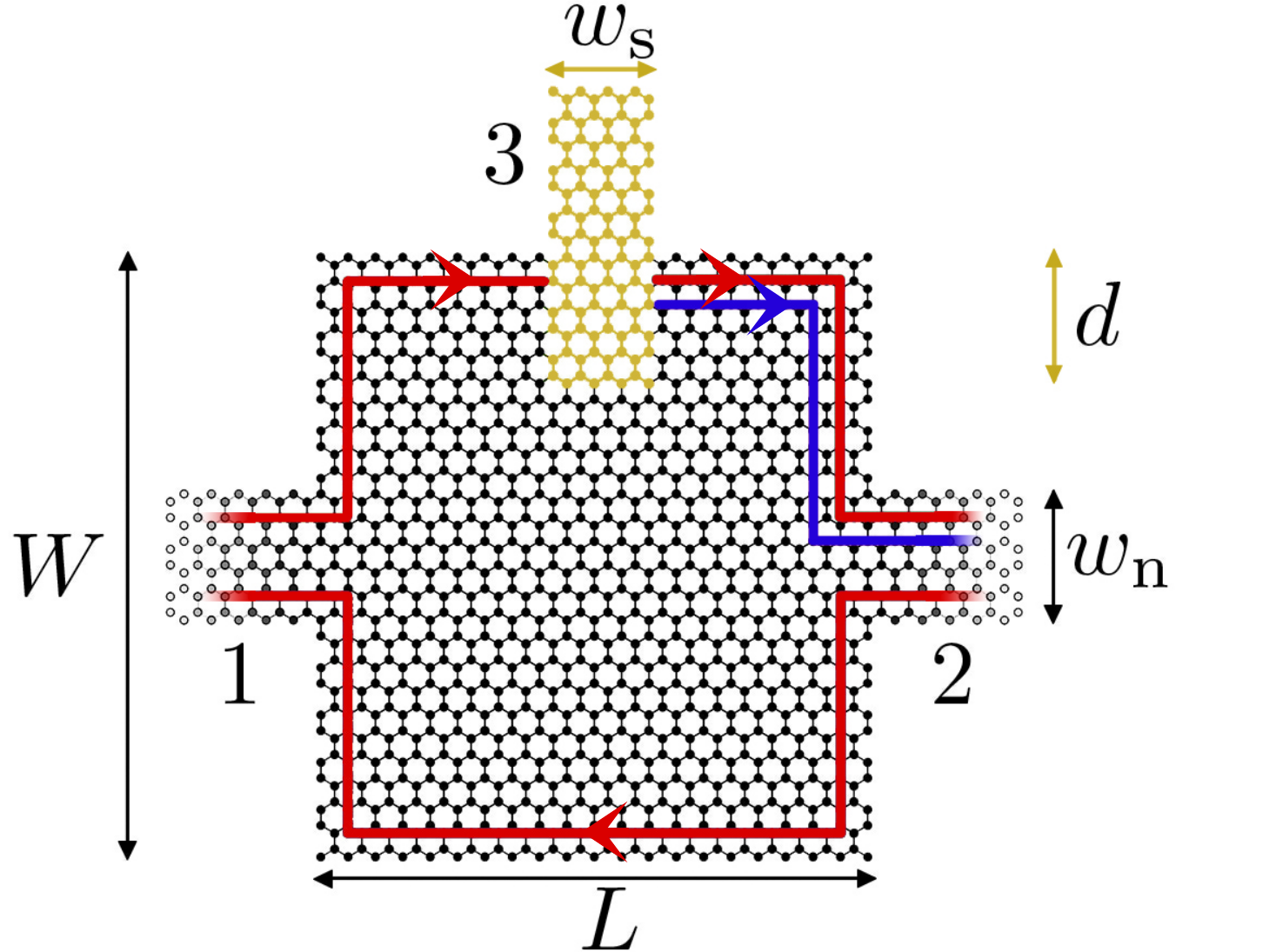}
\caption{(Color online) Pictorial representation of the three-terminal G/S setup with zigzag horizontal edges analysed in this work. The leads are labeled by numbers, while $W$ and $L$ represent, respectively, the width and the length of the GNR, $w_{\rm s}$ and $d$ the width and the penetration length in the normal region of the S finger, and $w_{\rm n}$ the width of the normal leads.
Clockwise oriented red (blue) thick lines represent the trajectories of electron (hole) QH edge states. After impinging on the S finger, electrons coming from the left (red line) are partially crossed Andreev reflected on the right as holes (blue line), and partially normal transmitted (red line).
\label{fig:one}}
\end{figure}

In this Article, we investigate the transport in the QH regime through a three-terminal G/S system resembling the structure experimentally studied in Ref.~\onlinecite{lee_natphys_2017} and sketched in Fig.~\ref{fig:one}.
We show that a negative nonlocal resistance arises when CAR exceeds normal transmission and we focus on the impact of lattice orientation on transport properties.
A rectangular GNR, of width $W$ and length $L$, is attached to two normal horizontal leads 1 and 2, of width $w_{\rm n}$, and to one superconducting vertical electrode 3, of width $w_{\rm s}$, that penetrates the normal region for a length $d$. The two normal leads and the superconducting one are attached, respectively, at half width and at half length.
We consider two opposite lattice orientations: the first one in which the horizontal edges are in the zigzag configuration (ZC) -- that is the lattice orientation depicted in Fig.~\ref{fig:one} -- and the second one in which the horizontal edges are in the armchair configuration (AC).
We find that the magnitude of normal transmission and CAR crucially depends on the lattice orientation of the GNR, the latter being favoured in the AC.

{\it Model Hamiltonian and resistances.---}In the tight-binding representation the graphene Hamiltonian reads
\begin{equation}\label{eq:graphene_hamiltonian}
	{\cal H} = -E_{\rm F} \sum_{i} c_{i}^\dagger c_{i} - \sum_{\langle ij\rangle} t_0 e^{i \phi_{ij}}c_{i}^\dagger c_{j} + {\rm h.c.}~,
\end{equation}
where $c_{i}$ and $c_{i}^\dagger$ are the annihilation and creation operators of a particle at the site $i$, $E_{\rm F}$ is the Fermi energy and $t_0$ is the nearest-neighbor hopping energy. The complex phase $\phi_{ij}$ is added in the hopping term using the Peierls substitution to take into account a uniform perpendicular magnetic field $B$ as $\phi_{ij} = 2\pi \phi_0^{-1} \int_i^j \bm A \cdot d\bm l$, described by a vector potential $\bm A = (-By, 0, 0)$ and $\phi_0=h/e$.
The presence of supercondcuting regions is accounted for by the Bogoliubov-de Gennes Hamiltonian~\cite{degennes_1989}
\begin{equation}
\label{bdg}
	{\cal H}_{\rm BdG} = 
	\begin{pmatrix}
		{\cal H} && \Delta \\
		\Delta^* && -{\cal H}^*
	\end{pmatrix}~,
\end{equation}
where the S (s-wave) order parameter, which couples electrons and holes, is a matrix $\Delta$ with entries $\Delta = \Delta_i \delta_{ij}$. Here $\Delta_i$ is taken to vary across the junction between its maximum value $\Delta_0$ in the bulk of the S and zero in the bulk of the GNR over a length $2a$, where $a$ is the lattice constant~\cite{rainis_prb_2009}. We assume the field $B$ to be absent in the S as a result of the Meissner effect.

The linear-response current-voltage relations for the normal leads are obtained within the Landauer-B\"uttiker scattering approach~\cite{datta_1997}. The current in lead $i$ is given by~\cite{lambertdag_jpcm_1998}
\begin{equation}
	I_i = \sum_{j=1}^2 a_{ij} (V_j - V)~,
\end{equation}
where $V_j$ is the voltage of the electrode $j$ and $V=\mu/e$, with $\mu$ the electrochemical potential of the superconducting condensate.
At zero temperature, the coefficients $a_{ij}$ read
\begin{equation}\label{eq:current_voltage_coefficients}
	a_{ij} = \frac{2e^2}{h} (N_j^{\rm e} \delta_{ij} - T_{ij}^{\rm ee} + T_{ij}^{\rm he})~,
\end{equation}
where the prefactor $2$ accounts for spin degeneracy and e (h) stands for electron (hole).
In Eq.~\eqref{eq:current_voltage_coefficients} $T_{ij}^{\rm ee}$ and $T_{ij}^{\rm he}$ are the normal and the Andreev scattering coefficients from lead $j$ to lead $i$ computed at the Fermi energy, respectively, and,
\begin{equation}\label{eq:open_channels}
	N_j^{\rm e} = \sum_{i}^{M} (T_{ij}^{\rm ee} + T_{ij}^{\rm he})
\end{equation}
is the number of open channels for electrons in lead $j$.

Similarly to the measurement configuration of Ref.~\onlinecite{lee_natphys_2017}, we impose a current bias between the normal lead 1 and the superconducting lead 3, while lead 2 is floating, i.e., $I_1 = I$ and $I_2 = 0$. We define the resistances $R_1 = (V_1-V)/I$ and $R_2 = (V_2-V)/I$, finding
\begin{equation}
	R_1 = \frac{h}{2e^2}\frac{T_{22}^{\rm he} + (T_{12}^{\rm he} + T_{12}^{\rm ee})/2}{D} \label{eq:nonlocal_res_1}
\end{equation}
and
\begin{equation}
	R_2 =  \frac{h}{2e^2} \frac{T_{21}^{\rm ee} - T_{21}^{\rm he}}{2D}~,\label{eq:nonlocal_res_2}
\end{equation}
where
\begin{multline}
	D = T_{12}^{\rm he}T_{21}^{\rm ee} + T_{12}^{\rm ee}T_{21}^{\rm he} + T_{11}^{\rm he} (T_{22}^{\rm he} + T_{12}^{\rm he} + T_{12}^{\rm ee}) + \\
	T_{22}^{\rm he} (T_{11}^{\rm he} + T_{21}^{\rm he} + T_{21}^{\rm ee})~.
\end{multline}
It is evident from Eqs.~\eqref{eq:nonlocal_res_1} and \eqref{eq:nonlocal_res_2} that the resistance $R_1$ is always positive, whereas the {\it nonlocal} resistance $R_2$ becomes negative if the CAR $T_{21}^{\rm he} \equiv T^{\rm A}$ is greater than the normal transmission $T_{21}^{\rm ee} \equiv T^{\rm N}$.


{\it Results.---}The transmission coefficients $T_{ij}^{\alpha\beta}$ are numerically calculated using Kwant~\cite{groth_njp_2014}, a toolkit which implements a wave function matching technique.
Unless otherwise stated, we use the following parameters: $L=100.0$ nm, $W=116.8$ nm, $w_{\rm n} = 5.4$ nm, $d=28.5$ nm (with $a= 0.246$ nm), $t_0=2.8$ eV, and we fix $\Delta_0 = 40$ meV and $B = 200$ Tesla in order to obtain a ratio between the magnetic length $\ell_B$ and the superconducting coherence length $\xi_{\rm s}$ of the same order as for the experiment in Ref.~\onlinecite{lee_natphys_2017}.
In particular, we have $\xi_{\rm s} = \hbar v_{\rm F}/(\pi \Delta_0) \simeq 4.75$ nm and $\ell_B = \sqrt{\phi_0/(2\pi B)} \simeq 1.84$ nm, so that the ratio $\ell_B/\xi_{\rm s} \simeq 0.39$.
Note that the number of open channels for electrons in the two leads is equal and given by the filling factor $\nu$ ($N^{\rm e}_1 =N^{\rm e}_2=\nu$), which is related to the Landau level index $N$ through $\nu=2N+1$.~\cite{goerbig_rmp_2011}

\begin{figure}[t!]
\centering
\begin{overpic}[width=.49\linewidth]{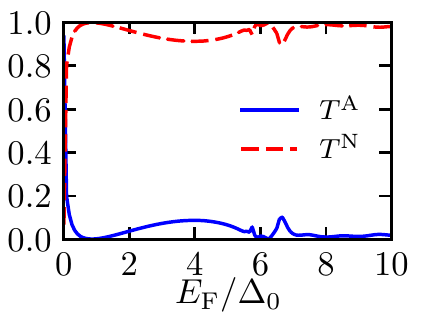}\put(0,77){(a)}\end{overpic}
\begin{overpic}[width=.49\linewidth]{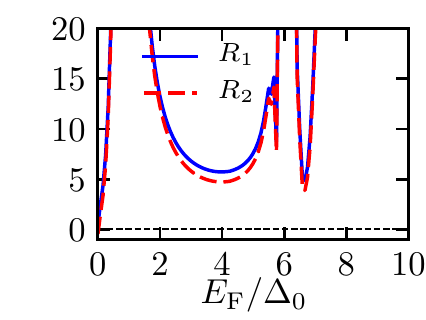}\put(0,77){(b)}\end{overpic}
\begin{overpic}[width=.49\linewidth]{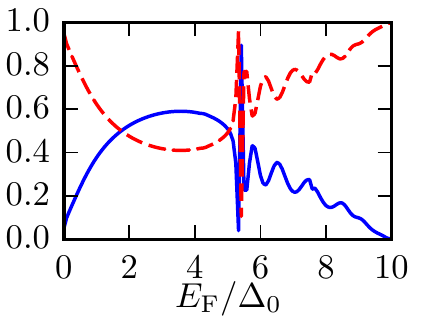}\put(0,77){(c)}\end{overpic}
\begin{overpic}[width=.49\linewidth]{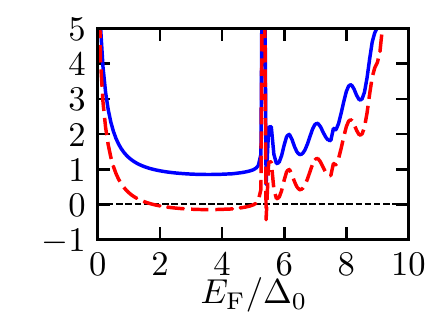}\put(0,77){(d)}\end{overpic}
\caption{(Color online) (a)-(b)~Narrow finger in the ZC. Numerical results for (a)~transmission coefficients [CAR $T^{\rm A}$ (solid blue) and normal transmission $T^{\rm N}$ (dashed dotted red)], and for (b)~resistances [$R_1$ (solid blue) and $R_2$ (dashed red)], in units of $h/(2e^2)$.
(c)-(d)~Narrow finger in the AC. Same convention as for panels (a) and (b).
The geometrical dimensions are: $L = 100.0$ nm, $W = 116.8$ nm, $d = 28.5$ nm, $w_{\rm s} = 5.4$ nm, $w_{\rm n} = 19.9$ nm. The tight-binding parameters  are: $\Delta_0 = 40.0$ meV, $t_0=2.8$ eV, and $B = 200$ T, so that the ratio $\ell_B/\xi_{\rm s} \approx 0.39$ is close to the experimental value.
\label{fig:two}}
\end{figure}

Let us first consider the case in which the width of the S is of the same order of the superconducting coherence length (narrow finger, $w_{\rm s} = 5.4$ nm) and focus on the range of values of $E_{\rm F}$ between 0 and $10\Delta_0$, where only the lowest Landau level ($N=0$) contributes to transport.
In Fig.~\ref{fig:two} we show the main numerical results obtained for the system in the ZC [panels (a) and (b)] and in the AC [panels (c) and (d)].
In particular, in Fig.~\ref{fig:two}(a) we show normal transmission and CAR ($T^{\rm N}$ and $T^{\rm A}$, respectively) as functions of the Fermi energy $E_{\rm F}$.
By increasing $E_{\rm F}$ from zero, $T^{\rm A}$ varies quite smoothly up to $E_{\rm F}/\Delta_0\simeq 5$ where a few peaks appear (whose origin is beyond the scope of this paper).

Interestingly, $T^{\rm A}$ remains much smaller than $T^{\rm N}$ for all values of $E_{\rm F}>0$.
We have verified that this behavior is not sensitive to the actual value of the ratio $\ell_B/\xi_{\rm s}$ (we have checked several values ranging from about 0.1 to 2).
Note that the relation $T^{\rm N} + T^{\rm A} = N^{\rm e}_2 =1$ holds as a consequence of Eq.~\eqref{eq:open_channels} and of the fact that all the reflections coefficients are zero (i.e., $T_{ii}^{\rm ee}= T_{ii}^{\rm he}=0$ with $i=1,2,3$) because of the chiral nature of the edge states~\cite{hou_prb_2016}.
Moreover, the normal transmission between lead 2 and 1, $T_{12}^{\rm ee}$, is exactly 1 since the QH edge state on the lower side of the Hall bar does not encounter any superconducting region to convert electrons into holes (see Fig.~\ref{fig:one}).

In Fig.~\ref{fig:two}(b) the resistances $R_1$ and $R_2$ are shown as functions of the Fermi energy.
As a consequence of the small value of $T^{\rm A}$, such resistances are very large and the nonlocal resistance $R_2$ is always positive.
This is made clear by the following expressions
\begin{equation}
	R_1 = \frac{h}{2e^2} \frac{1}{2 T^{\rm A}}
	\label{eq:nonlocal_res_1_QH}
\end{equation}
and
\begin{equation}
	R_2 = \frac{h}{2e^2} \left( \frac{1}{2 T^{\rm A}} - \frac{1}{\nu} \right) ,
	\label{eq:nonlocal_res_2_QH}
\end{equation}
obtained as a result of the fact that all reflection coefficients vanish and using Eqs.~\eqref{eq:open_channels}, \eqref{eq:nonlocal_res_1}, and \eqref{eq:nonlocal_res_2} (the number of open channels in leads 1 and 2 are assumed to be equal).
It is worth pointing out that Eqs.~\eqref{eq:nonlocal_res_1_QH} and \eqref{eq:nonlocal_res_2_QH} still hold true when additional floating normal terminals are included in the setup.
Moreover, the difference $R_1-R_2$ takes the constant value $h/(2e^2\nu)$, representing the Hall resistance.

\begin{figure}[t!]
\centering
\begin{overpic}[width=.49\linewidth]{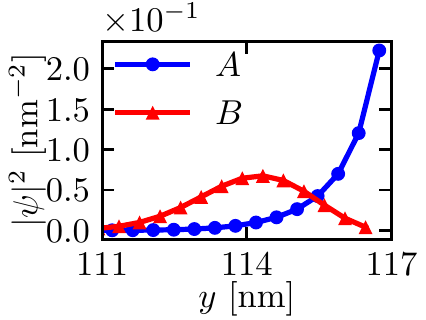}\put(0,77){(a)}\end{overpic}
\begin{overpic}[width=.49\linewidth]{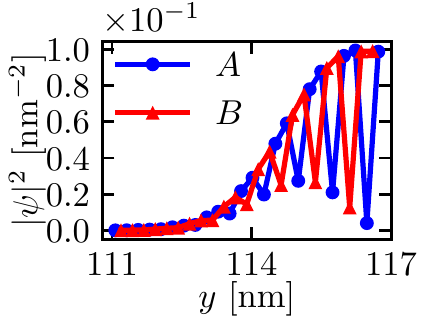}\put(0,77){(b)}\end{overpic}
\caption{(Color online) 
Square modulus of the electron wave function for $E_{\rm F} = 4 \Delta_0$ along the $y$-direction for the ZC (a) and AC (b), close to the top edge of graphene [panel (b) is a vertical cut of Fig.~\ref{fig:four}(a) taken at at $x\simeq 14$ nm]. All other parameters are the same as for Fig.~\ref{fig:two}.\label{fig:three}}
\end{figure}

Let us now consider the AC.
Contrary to the ZC case, Fig.~\ref{fig:two}(c) shows that $T^{\rm A}$ takes a small value at $E_{\rm F}$ close to zero, and thereafter smoothly increases taking values larger than $T^{\rm N}$ in a wide range of Fermi energies.
Sharp features appear for $E_{\rm F}/\Delta_0>5$ as $T^{\rm A}$ exhibits an overall decrease.
Also for the AC, this behavior is not sensitive to the ratio $\ell_B/\xi_{\rm s}$.
As shown in Fig.~\ref{fig:two}(d), in correspondence to the values of $E_{\rm F}$ for which $T^{\rm A}$ becomes larger than $T^{\rm N}$, the nonlocal resistance $R_2$ becomes negative.
This suggests that the lattice orientation of the GNR is crucial as far as the sign of the nonlocal resistance $R_2$ is concerned for narrow S fingers.
The role of the lattice orientation on the behavior of $T^{\rm A}$ for edge states in the lowest Landau level could be imputed to the following two facts.
(i) The spatial distribution of the wave functions on the two sublattices ($A$ and $B$) are different in the ZC, while they are very close to each other~\cite{brey_prb_2006} (same envelop function) in the AC [see Figs.~\ref{fig:three}(a) and (b), respectively]; (ii) In the ZC the wave function of valley K (K$'$) is localized on sublattice $A$ ($B$), i.e., valley and sublattice indices are locked, while in the AC the wave function is spread over the two sublattices~\cite{ACZC}.
Now, electrons and holes involved in the Andreev processes belong to different valleys. Therefore, $T^{\rm A}$ is unfavored in the ZC since, unlike in the AC, a valley switch requires the wavefunction to change sublattice (with a different spatial distribution).

In both ZC and AC cases, the smooth variation of the transmission coefficients, as functions of the Fermi energy, originates from the fact that particle transfer takes place through the bulk of the S narrow finger via normal transmission and CAR.
Indeed, the space-resolved square modulus of the particle (electron or hole) wave function turns out to be localized along the top horizontal edge without following the profile of the G/S interface.
This is pictorially shown in Fig.~\ref{fig:one} where the red (blue) lines represent the electron (hole) QH edge states.
The behavior of the transmission coefficients as functions of the Fermi energy changes dramatically in the case of a wide finger (where the width of S is much larger than the superconducting coherence length).
In this situation both $T^{\rm A}$ and $T^{\rm N}$ vary rapidly with $E_{\rm F}$, exhibiting fast fluctuations.~\cite{TA0}
In particular, in the AC, $T^{\rm A}$ and $T^{\rm N}$ fluctuations intersect each other, so that $T^{\rm A}$ is very often larger than $T^{\rm N}$ (and $R_2$ is negative).
In the ZC, however, the fluctuation amplitudes of $T^{\rm A}$ are generally small so that $T^{\rm N}$ is larger than $T^{\rm A}$ in a wide range of values of $E_{\rm F}$ (and $R_2$ is positive).

\begin{figure}[t]
\centering
\begin{overpic}[width=1\linewidth]{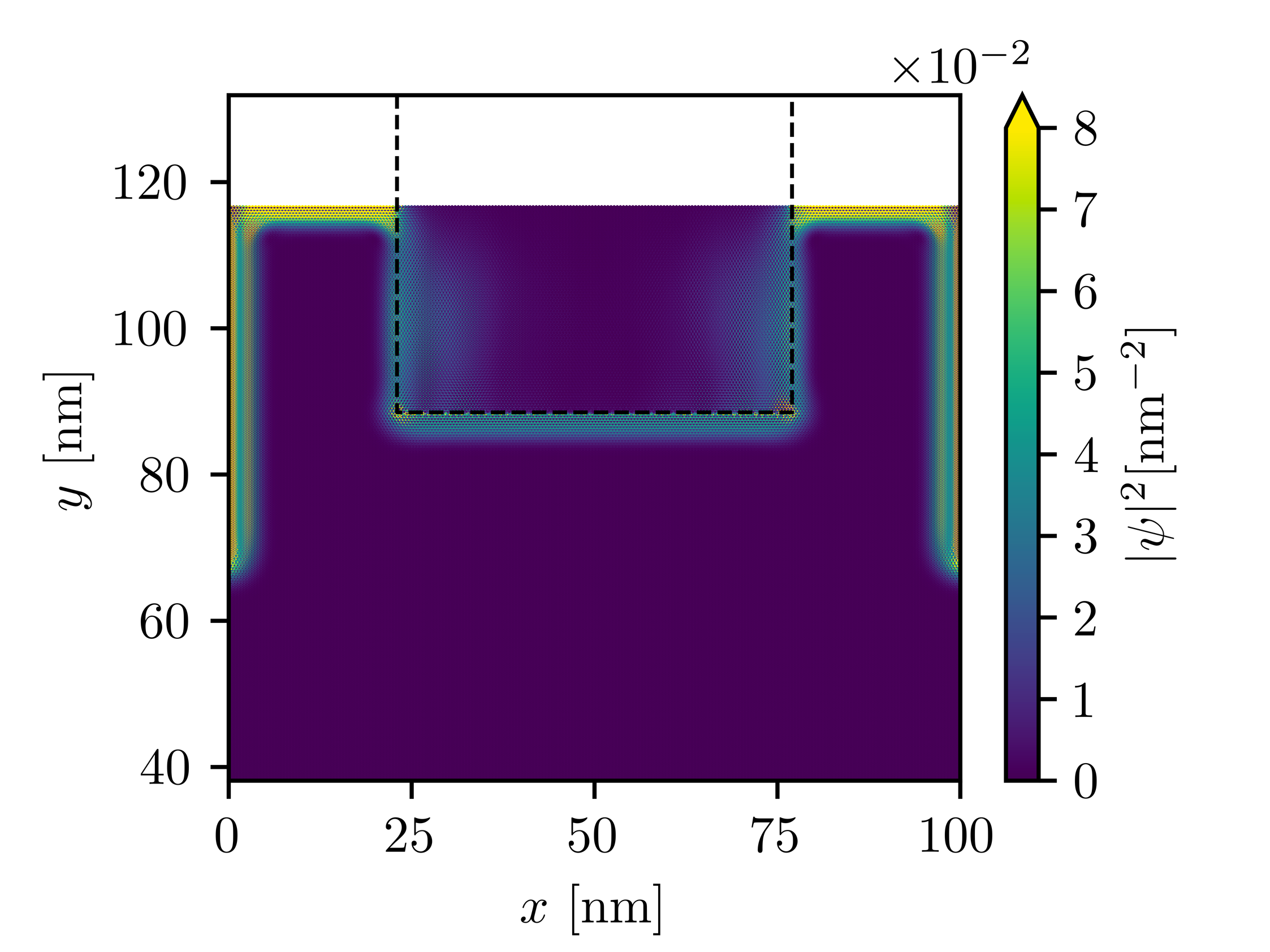}\put(0,70){(a)}\end{overpic}
\begin{overpic}[width=1\linewidth]{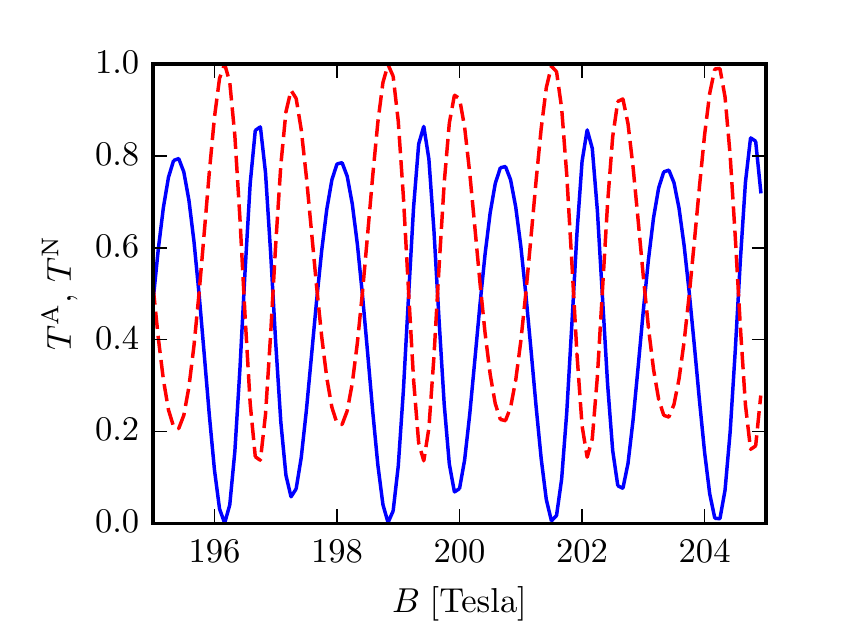}\put(0,70){(b)}\end{overpic}
\caption{(Color online) Numerical results for (a)~the square modulus of the electron wave function determined by electrons incoming from lead $1$ and (b)~the transmission coefficients [CAR $T^{\rm A}$ (solid blue) and normal transmission $T^{\rm N}$ (dashed red)] versus the magnetic field for the AC and $w_{\rm s} = 54.0$ nm at $E_{\rm F} = 4 \Delta_0$. All the other parameters are the same as for Fig.~\ref{fig:two}. The magnetic field $B$ ranges from $195$ to $205$ T, so that the ratio $\ell_B/\xi_{\rm s}$ remains close to the experimental values. \label{fig:four}}
\end{figure}

Let us now discuss more in details the case of a wide finger ($w_{\rm s} = 54.0$ nm), focusing on the system in the AC.
In Fig.~\ref{fig:four}(a) we plot the square modulus $|\psi|^2$ of the electron component of the wavefunction which satisfies the Bogoliubov-de Gennes equation for the Hamiltonian~(\ref{bdg}), resulting from electrons incoming from lead 1 (centered at $y \simeq 58.4$ nm).
In Fig.~\ref{fig:four}(a) the location of the S finger is marked by a black dashed line (note that the upper half of the setup only is represented).
The wavefunction represents electrons following the edge of the GNR by propagating clockwise as an edge state with electron component only.
At $x\sim 20$ nm on the top edge, the electrons impinge on the S finger and the edge state follows the G/S interface up to $x\sim 80$ nm on the top edge, where the edge state leaves the S finger, finally ending up in lead 2.
Along the G/S interface the edge state is a mixture of electrons and holes as a consequence of the Andreev processes occurring at the interface (note that in Fig.~\ref{fig:four}(a) the electron component only is plotted).

Contrary to what happens for a narrow finger, here transport occurs along the G/S interface though such Andreev edge states.
Unexpectedly, however, the wave function horizontally penetrates for a rather long distance in the S finger (about 3 times $\ell_B$ and decaying with a few oscillations), while the wave function decay is abrupt in the vertical direction.
It is worth mentioning that $|\psi|^2$ turns out to be peaked at the two corners of the G/S interface. 

Fig.~\ref{fig:four}(b), where the transmission coefficients are plotted as functions of the magnetic field $B$, shows that $T^{\rm N}$ and $T^{\rm A}$ exhibit wide and regular oscillations (note that $T^{\rm A}$ ranges from 0 to 0.9, so that $T^{\rm A}$ always crosses $T^{\rm N}$).
They originate from interference effects between electron and hole components of the Andreev edge states (see also Refs.~\onlinecite{hoppe_prl_2000,giazotto_prb_2005,khaymovich_epl_2010}).
Interestingly, the period of these oscillations is independent of the Fermi energy.
A similar behaviour is observed in the ZC, but generally the oscillations of $T^{\rm A}$ are of much smaller amplitude so that $T^{\rm N}$ remains larger than $T^{\rm A}$.
In the AC, therefore, the magnetic field can be used as a mean to tune $T^{\rm A}$ and change the sign of $R_2$.~\cite{narrowB}

{\it Conclusions.---}In this Article we have investigated Andreev processes, local, and nonlocal resistances in a GNR where a superconducting finger is inserted.
We have considered two different situations, namely a narrow and a wide finger (compared with the superconducting coherence length).
In both cases, we have found that the relative magnitude of CAR versus normal transmission is very sensitive to the lattice orientation.
In particular, CAR is generally dominant (over normal transmission) when the graphene edge at the G/S interface is armchair, and results in a negative nonlocal resistance.
Moreover, while transmission through a narrow finger occurs via tunnelling, transport with a wide finger occurs through Andreev edge states that develop at the G/S interface and allow a magnetic field tuning of CAR.

We believe that our results enable a deeper understanding of the measurements reported in Ref.~\onlinecite{lee_natphys_2017}.
Indeed, we have demonstrated that in the QH regime the negative nonlocal resistance is clearly associated to the Andreev transmission being larger than the normal transmission. Moreover, in the case of narrow S, the edge-sensitive behaviour of the Andreev transmission could be used to distinguish between ZC and AC cases.

{\it Acknowledgments.---}This work was supported by the European Union's Horizon 2020 research and innovation programme under grant agreement No. 696656 ``GrapheneCore'' and the SNS internal project ``Non-equilibrium dynamics of one-dimensional quantum systems: From synchronization to many-body localization''. F.T. acknowledges support from the SNS internal project ``Thermoelectricity in nanodevices'', the MIUR-QUANTRA and the CNR-CONICET cooperation programmes. Free software (www.gnu.org, www.python.org) was used.


\begin{thebibliography}{77}
\bibitem{narrowB}
For completeness we mention that for a narrow S finger the transmission coefficients are virtually independent of $B$.
\end{thebibliography}
\end{document}